*Research Article*

# The Unified Hydrodynamics and the Pseudorapidity Distributions in Heavy Ion Collisions at BNL-RHIC and CERN-LHC Energies

Z. J. Jiang, J. Wang, K. Ma, and H. L. Zhang

*College of Science, University of Shanghai for Science and Technology, Shanghai 200093, China*

Correspondence should be addressed to Z. J. Jiang; jzj265@163.com

The charged particles produced in nucleus-nucleus collisions are divided into two parts. One is from the hot and dense matter created in collisions. The other is from leading particles. The hot and dense matter is assumed to expand according to unified hydrodynamics and freezes out into charged particles from a space-like hypersurface with a fixed proper time of $\tau_{\text{FO}}$. The leading particles are conventionally taken as the particles which inherit the quantum numbers of colliding nucleons and carry off most of incident energy. The rapidity distributions of the charged particles from these two parts are formulated analytically, and a comparison is made between the theoretical results and the experimental measurements performed in Au-Au and Pb-Pb collisions at the respective BNL-RHIC and CERN-LHC energies. The theoretical results are well consistent with experimental data.

## 1. Introduction

Since the elementary work of Landau in 1953 [1], relativistic hydrodynamics has been applied to calculate a large number of variables developed in the context of nucleon or nucleus collisions at high energy. In particular, owing to the successful description of elliptic flow measured at the Relativistic Heavy-Ion Collider (RHIC) at Brookhaven National Laboratory (BNL) [2] and recently at the Large Hadron Collider (LHC) at CERN [3], the hydrodynamic research has entered into a more active phase. It has now been widely accepted as one of the best approaches for understanding the space-time evolution of the matter created in such collisions [4–10].

Although at present there are powerful numerical approaches to deal with certain hydrodynamic problems, this will require a very large scale of calculation and skillful sophisticated techniques to avoid instabilities. On the contrary, analytical solutions, since their simple and transparent forms, usually providing us with an invaluable insight into the characteristics of matter created in collisions, are always our pursuit of the goal though usually at the price of some ideal assumptions.

Due to the tremendous complexity of hydrodynamic equations, the progress in finding exact solutions is not going well. Up till now, most of this work mainly involves 1 + 1 dimensional flows with simple equation of state [11–19]. The 3 + 1 dimensional hydrodynamics is less developed, and no general exact solutions are known so far.

An important application of 1 + 1 dimensional hydrodynamics is the analysis of the pseudorapidity distributions of the charged particles produced in nucleon or nucleus collisions. In this paper, by taking into account the effect of leading particles, we will discuss such distributions in the context of unified hydrodynamics [14]. The main points of this model are listed in Section 2. Its solution is then exploited in Section 3 to formulate the rapidity distributions of the charged particles frozen out from fluid at a space-like hypersurface with a fixed proper time $\tau_{\text{FO}}$. In Section 4, the theoretical results are compared with the experimental data performed in nucleus-nucleus collisions at BNL-RHIC and CERN-LHC energies. Section 5 is about conclusions.

## 2. A Brief Description of the Unified Hydrodynamics

The 1 + 1 expansion of a perfect fluid follows the equations:

$$\frac{\partial T^{00}}{\partial t} + \frac{\partial T^{10}}{\partial z} = 0,$$
$$\frac{\partial T^{01}}{\partial t} + \frac{\partial T^{11}}{\partial z} = 0, \quad (1)$$

where $t$ is the time, $z$ is the longitudinal coordinate along beam direction, and

$$T^{\mu\nu} = (\varepsilon + p) u^\mu u^\nu - p g^{\mu\nu} \quad (2)$$

is the energy-momentum tensor, and $g^{\mu\nu} = \mathrm{diag}(1, -1, -1, -1)$, $u^\mu$, $\varepsilon$, and $p$ are, respectively, the metric tensor, 4-velocity, energy density, and pressure of fluid. For a constant speed of sound, $\varepsilon$ and $p$ are related by the equation of state:

$$\varepsilon = gp, \quad (3)$$

where $1/\sqrt{g} = c_s$ is the speed of sound. Investigations have shown that $g$ changes very slowly with energy and centrality [20–23]. For a given incident energy, it can be well taken as a constant.

Using (2) and (3), and noticing the light-cone components of the 4-velocity

$$u_\pm = e^{\pm y}, \quad (4)$$

where $y$ is the ordinary rapidity of fluid, (1) reads

$$\frac{e^{2y} - 1}{2}(g+1)\partial_+ p + e^{2y}(g+1)p\partial_+ y + \frac{1 - e^{-2y}}{2}(g+1)\partial_- p$$
$$+ e^{-2y}(g+1)p\partial_- y + \partial_+ p - \partial_- p = 0,$$
$$\frac{e^{2y} + 1}{2}(g+1)\partial_+ p + e^{2y}(g+1)p\partial_+ y + \frac{1 + e^{-2y}}{2}(g+1)\partial_- p$$
$$- e^{-2y}(g+1)p\partial_- y - \partial_+ p - \partial_- p = 0, \quad (5)$$

where $\partial_+$ and $\partial_-$ are the compact notation of partial derivatives with respect to light-cone coordinates $z_\pm = t \pm z = x^0 \pm x^1 = \tau e^{\pm \eta_{\mathrm{ST}}}$, $\tau = \sqrt{z_+ z_-}$ is the proper time, and $\eta_{\mathrm{ST}} = 1/2 \ln(z_+/z_-)$ is the space-time rapidity of fluid. The solutions of above equations are

$$g\partial_+ \ln p = -\frac{(g+1)^2}{2}\partial_+ y - \frac{g^2 - 1}{2}e^{-2y}\partial_- y,$$
$$g\partial_- \ln p = \frac{(g+1)^2}{2}\partial_- y + \frac{g^2 - 1}{2}e^{2y}\partial_+ y. \quad (6)$$

The key ingredient of unified hydrodynamics is that it generalizes the relation between $y$ and $\eta_{\mathrm{ST}}$ by

$$2y = \ln u_+ - \ln u_- = \ln F_+(z_+) - \ln F_-(z_-), \quad (7)$$

where $F_\pm(z_\pm)$ are a priori arbitrary function obeying equation

$$F_\pm F''_\pm = \frac{A^2}{2}, \quad (8)$$

where $A$ is a constant. In case of

$$F_\pm(z_\pm) = z_\pm. \quad (9)$$

Equation (7) reduces to $y = \eta_{\mathrm{ST}}$, returning to the boost-invariant picture of Hwa-Bjorken. Otherwise, (7) describes the nonboost-invariant geometry of Landau. Accordingly, (7) unifies the Hwa-Bjorken and Landau hydrodynamics together. It paves a way between these two models.

Substituting (7) into (6), we have

$$g\partial_+ \ln p = -\frac{(g+1)^2}{4}\frac{f'_+}{f_+} + \frac{g^2 - 1}{4}\frac{f'_-}{f_+},$$
$$g\partial_- \ln p = -\frac{(g+1)^2}{4}\frac{f'_-}{f_-} + \frac{g^2 - 1}{4}\frac{f'_+}{f_-}, \quad (10)$$

where $f_\pm = F_\pm/H$ and $H$ is an arbitrary constant. From the above equations, we can get the entropy density of fluid:

$$s(z_+, z_-) = s_0 \left(\frac{p}{p_0}\right)^{g/(g+1)}$$
$$= s_0 \exp\left[-\frac{g+1}{4}(l_+^2 + l_-^2) + \frac{g-1}{2}l_+ l_-\right], \quad (11)$$

where, by definition

$$l_\pm(z_\pm) = \sqrt{\ln f_\pm}, \quad (12)$$

and in terms of $l_\pm$, we have from (7) and (8)

$$y(z_+, z_-) = \frac{1}{2}(l_+^2 - l_-^2), \quad (13)$$

$$z_\pm = 2h \int_0^{l_\pm} e^{u^2} du, \quad (14)$$

where $h = H/A$.

## 3. The Rapidity Distributions of the Charged Particles Frozen out from Fluid

By using (11), we can obtain the rapidity distributions of the charged particles frozen out from fluid or hot and dense matter created in collisions. To this end, we first evaluate the entropy distributions of the fluid on a space-like hypersurface with a fixed proper time $\tau_{\mathrm{FO}}$, from which the fluid will freeze out into the charged particles. Such distributions take the form

$$\frac{dS}{dy} = s u^\mu \frac{d\lambda_\mu}{dy}\bigg|_{\tau_{\mathrm{FO}}} = s u^\mu n_\mu \frac{d\lambda}{dy}\bigg|_{\tau_{\mathrm{FO}}}, \quad (15)$$

where $n^\mu$ is the 4-dimensional unit vector normal to the hypersurface

$$n^\mu n_\mu = n_+ n_- = 1. \qquad (16)$$

Also $d\lambda^\mu = d\lambda n^\mu$, and $d\lambda$ is the space-like infinitesimal length element along hypersurface

$$d\lambda = \sqrt{d\lambda^\mu d\lambda_\mu} = \sqrt{-dz^+ dz^-}. \qquad (17)$$

Considering a hypersurface

$$\phi(z_+, z_-) = C, \qquad (18)$$

where $C$ is a constant, we have

$$\partial_+\phi dz_+ + \partial_-\phi dz_- = 0. \qquad (19)$$

The light-cone components of unit vector $n^\mu$ can be expressed by $\partial_\pm \phi$ as

$$n_+ = \frac{\partial_-\phi}{\sqrt{\partial_+\phi \partial_-\phi}}, \quad n_- = \frac{\partial_+\phi}{\sqrt{\partial_+\phi \partial_-\phi}},$$

$$d\lambda = \sqrt{-dz_+ dz_-} = dz_-\sqrt{\frac{\partial_-\phi}{\partial_+\phi}} = dz_+\sqrt{\frac{\partial_+\phi}{\partial_-\phi}}. \qquad (20)$$

Then, the expression on the right-hand side of (15) is

$$u^\mu n_\mu d\lambda = \frac{1}{2}(u_+ n_- + u_- n_+) d\lambda$$

$$= \frac{1}{2}(u_+\partial_+\phi + u_-\partial_-\phi)\frac{dz_-}{\partial_+\phi} \qquad (21)$$

$$= \frac{1}{2}(u_+\partial_+\phi + u_-\partial_-\phi)\frac{dz_+}{\partial_-\phi}.$$

Furthermore, known from (14),

$$dz_\pm = \frac{h}{l_\pm}\exp(l_\pm^2) dl_\pm^2. \qquad (22)$$

In terms of $\partial_\pm\phi$ and $l_\pm$, $dl_\pm^2$ can be written as

$$dl_\pm^2 = \pm \frac{2\partial_\mp\phi \exp(l_\mp^2)/l_\mp}{\partial_+\phi \exp(l_+^2)/l_+ + \partial_-\phi \exp(l_-^2)/l_-} dy. \qquad (23)$$

Thus, (22) becomes

$$dz_\pm = \pm \frac{2h\partial_\mp\phi}{\partial_+\phi l_- \exp(-l_-^2) + \partial_-\phi l_+ \exp(-l_+^2)} dy. \qquad (24)$$

Substituting it into (21) and then into (15), we have

$$\frac{dS}{dy} = s e^{(1/2)(l_+^2+l_-^2)} \frac{\partial_+\phi e^y + \partial_-\phi e^{-y}}{\partial_+\phi l_- e^y + \partial_-\phi l_+ e^{-y}}\bigg|_{\tau_{FO}}. \qquad (25)$$

In the above equation, the right-hand side is evaluated at the hypersurface with proper time $\tau_{FO}$. Known from (18), this hypersurface can be taken as

$$\phi(z_+, z_-) = \tau_{FO}^2 = z_+ z_- = C, \quad z_+ dz_- + z_- dz_+ = 0. \qquad (26)$$

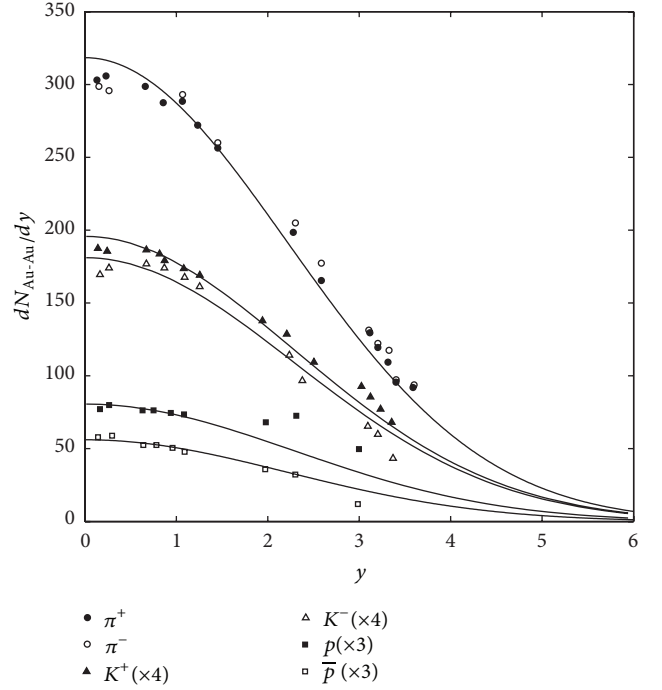

FIGURE 1: The rapidity distributions of the specified charged particles in central Au-Au collisions at $\sqrt{s_{NN}} = 200$ GeV. The scattered symbols are the experimental measurements [24–26]. The solid curves are the results from unified hydrodynamics of (28).

Comparing the above equation with (19), we get

$$\partial_\pm\phi = z_\mp. \qquad (27)$$

Inserting it together with (11) into (25) and noticing the proportional relation between entropy and the number of the charged particles, we have

$$\frac{dN_{Fluid}(b, \sqrt{s_{NN}}, y)}{dy}$$

$$= C(b, \sqrt{s_{NN}}) e^{-(g-1)(l_+-l_-)^2/4} \frac{z_- e^y + z_+ e^{-y}}{z_- l_- e^y + z_+ l_+ e^{-y}}, \qquad (28)$$

where $C(b, \sqrt{s_{NN}})$, independent of rapidity $y$, is an overall normalization constant. $b$ is the impact parameter, and $\sqrt{s_{NN}}$ is the center-of-mass energy per pair of nucleons.

## 4. Comparison with Experimental Measurements and the Rapidity Distributions of Leading Particles

From (28), we can get the rapidity distributions of the charged particles generated in collisions. Figure 1 presents such distributions for $\pi^+$, $\pi^-$, $K^+$, $K^-$, $p$, and $\bar{p}$ resulting from 5% most central Au-Au collisions at $\sqrt{s_{NN}} = 200$ GeV. The symbols are the experimental measurements given by BRAHMS Collaboration at BNL-RHIC [24–26]. The solid curves are the theoretical predictions from (28). From this figure, we can see that (28) coincides well with the experimental data of all the charged particles with the exception of $p$. For proton $p$, an obvious discrepancy appears in the rapidity interval of $y = 2.0 \sim 3.0$. This might be caused by the effects of leading particles.

The leading particles are believed to be formed outside the nucleus, that is, outside the colliding region [27, 28]. The generation of leading particles is therefore free from fluid evolution. Accordingly, their rapidity distributions are beyond the scope of hydrodynamics and should be treated separately.

In our previous work [6], we once argued that the rapidity distribution of leading particles takes the Gaussian form:

$$\frac{dN_{\text{Lead}}(b, \sqrt{s_{\text{NN}}}, y)}{dy} = \frac{N_{\text{Lead}}(b, \sqrt{s_{\text{NN}}})}{\sqrt{2\pi}\sigma} \exp\left\{-\frac{[|y| - y_0(b, \sqrt{s_{\text{NN}}})]^2}{2\sigma^2}\right\}, \quad (29)$$

where $N_{\text{Lead}}(b, \sqrt{s_{\text{NN}}})$, $y_0(b, \sqrt{s_{\text{NN}}})$, and $\sigma$ are, respectively, the number of leading particles, central position, and width of distribution. This conclusion comes from the consideration that, for a given incident energy, different leading particles resulting from each time of nucleus-nucleus collisions have approximately the same amount of energy or rapidity. Then, the central limit theorem [29, 30] guarantees the rationality of above argument. Actually, as one can see from the shapes of the curves in Figure 1 any kind of charged particles resulting in collisions forms a good Gaussian rapidity distribution.

Term $y_0(b, \sqrt{s_{\text{NN}}})$ in (29) is the average position of leading particles. It should increase with incident energy and centrality. The value of $\sigma$ relies on the relative energy or rapidity differences among leading particles. It should not, at least not apparently, depend on the incident energy, centrality, and even colliding system. The concrete values of $y_0$ and $\sigma$ can be determined by tuning the theoretical predictions to experimental data.

By definition, leading particles mean the particles which carry on the quantum numbers of colliding nucleons and take away most of incident energy. Then, the number of leading particles is equal to that of participants. For nucleon-nucleon, such as $p$-$p$ collisions, there are only two leading particles. They are separately in projectile and target fragmentation region. For an identical nucleus-nucleus collision, the number of leading particles is

$$N_{\text{Lead}}(b, \sqrt{s_{\text{NN}}}) = \frac{N_{\text{Part}}(b, \sqrt{s_{\text{NN}}})}{2}, \quad (30)$$

where $N_{\text{Part}}(b, \sqrt{s_{\text{NN}}})$ is the total number of participants equaling [31, 32]

$$N_{\text{Part}}(b, \sqrt{s_{\text{NN}}}) = \int n_{\text{Part}}(b, \sqrt{s_{\text{NN}}}, s) \, d^2s, \quad (31)$$

where the integral variable $s$ is the transverse coordinates in the almond-shaped colliding region with respect to the center of one nucleus. $n_{\text{Part}}(b, \sqrt{s_{\text{NN}}}, s)$ is the total number of participants in the flux tube with a unit bottom area, located at position $s$ along beam direction. It takes the form

$$n_{\text{Part}}(b, \sqrt{s_{\text{NN}}}, s)$$
$$= T_A(s)\left\{1 - \exp\left[-\sigma_{\text{NN}}^{\text{in}}(\sqrt{s_{\text{NN}}}) T_B(s-b)\right]\right\} \quad (32)$$
$$+ T_B(s-b)\left\{1 - \exp\left[-\sigma_{\text{NN}}^{\text{in}}(\sqrt{s_{\text{NN}}}) T_A(s)\right]\right\},$$

where $\sigma_{\text{NN}}^{\text{in}}(\sqrt{s_{\text{NN}}})$, the inelastic nucleon-nucleon cross section, increases slowly with energy. For instance, for $\sqrt{s_{\text{NN}}} = 200$, $130$, $62.4$ GeV and $2.76$ TeV, $\sigma_{\text{NN}}^{\text{in}} = 42$, $41$, $36$, and $64 \pm 5$ mb [33, 34], respectively. The subscripts $A$ and $B$ in the above equation represent the projectile and target nucleus. $T_A(s)$ or $T_B(s-b)$ is the nuclear thickness function with the value equaling the nucleon number in the flux as defined above. It is equal to

$$T(s) = \int \rho(s, z) \, dz, \quad (33)$$

where

$$\rho(r) = \frac{\rho_0}{1 + \exp\left[(r - r_0)/a\right]} \quad (34)$$

is the well-known Woods-Saxon distribution of nuclear density. $a$ and $r_0$, taking somewhat different values in different papers [31], are, respectively, the skin depth and radius of nucleus. In this paper, we take the value of $a = 0.54$ fm and $r_0 = 1.12 A^{1/3} - 0.86 A^{-1/3}$ fm, where $A$ is the mass number of nucleus. Here, $\rho_0$ is another constant determined by condition

$$\int \rho(r) \, dV = A. \quad (35)$$

Known from the investigations in [6], (31) can give a correct result in different nucleus-nucleus collisions at different centrality and energy from BNL-RHIC to CERN-LHC scales.

From rapidity distributions, we can get pseudorapidity distributions by relation [35]

$$\frac{dN(b, \sqrt{s_{\text{NN}}}, \eta)}{d\eta} = \sqrt{1 - \frac{m^2}{m_T^2 \cosh^2 y}} \frac{dN(b, \sqrt{s_{\text{NN}}}, y)}{dy}, \quad (36)$$

where $m_T = \sqrt{m^2 + p_T^2}$ is the transverse mass and $p_T$ is the transverse momentum. The first factor on the right-hand side of the above equation is actually the Jacobian determinant. This transformation is closed by another relation:

$$y = \frac{1}{2} \ln\left[\frac{\sqrt{p_T^2 \cosh^2 \eta + m^2} + p_T \sinh \eta}{\sqrt{p_T^2 \cosh^2 \eta + m^2} - p_T \sinh \eta}\right]. \quad (37)$$

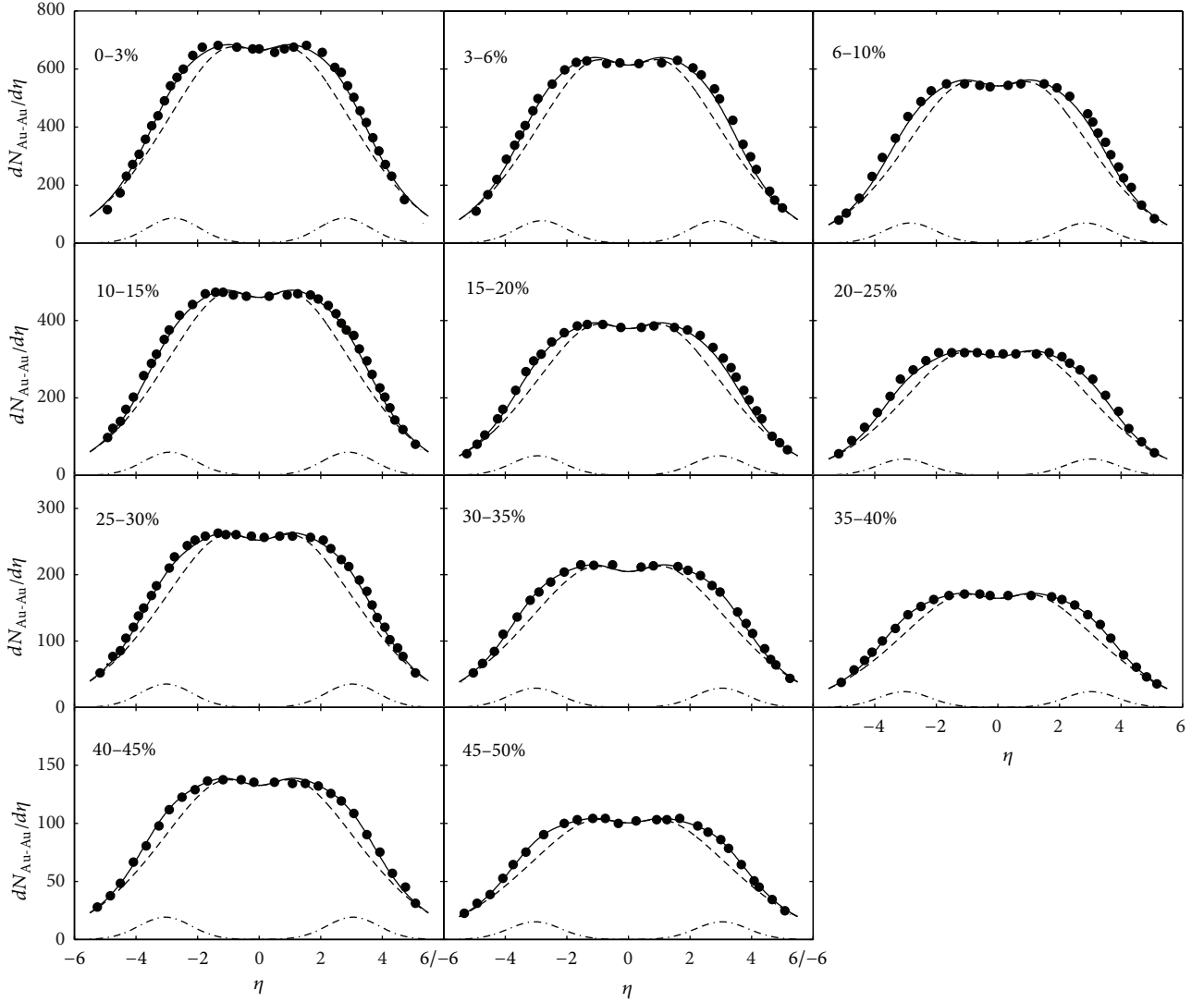

FIGURE 2: The pseudorapidity distributions of the charged particles produced in different centrality of Au-Au collisions at $\sqrt{s_{NN}}$ = 200 GeV. The solid dots are the experimental measurements [36]. The dashed curves are the results from unified hydrodynamics of (28). The dashed-dotted curves are the results from leading particles of (29). The solid curves are the sums of the dashed and dashed-dotted ones.

Taking into account the contributions from both the freeze-out of fluid and leading particles, the rapidity distributions in (36) can be written as

$$\frac{dN(b, \sqrt{s_{NN}}, y)}{dy} = \frac{dN_{Fluid}(b, \sqrt{s_{NN}}, y)}{dy} + \frac{dN_{Lead}(b, \sqrt{s_{NN}}, y)}{dy}. \quad (38)$$

Inserting the above equation or the sum of (28) and (29) into (36), we can get the pseudorapidity distributions of the charged particles. The results are shown in Figures 2, 3, 4, and 5, which are, respectively, for distributions in different centrality of Au-Au collisions at $\sqrt{s_{NN}}$ = 200, 130, and 62.4 GeV and in different centrality of Pb-Pb collisions at $\sqrt{s_{NN}}$ = 2.76 TeV. The solid dots are the experimental data [36, 37]. The dashed curves are the results obtained from unified hydrodynamics of (28). The dashed-dotted curves are the results got from leading particles of (29). The solid curves are the results acquired from (38), that is, the sums of the dashed and dashed-dotted curves. It can be seen that the theoretical results are in good agreement with experimental measurements.

In calculations, the parameter $g$ in (28) takes the values of $g$ = 8.16 and 5.55 in the respective Au-Au and Pb-Pb collisions at all of the concerned incident energy and centrality [20]. The width parameter $\sigma$ in (29) takes the value of 0.85 and 0.90 in Au-Au and Pb-Pb collisions, respectively, at different incident energy and centrality. As the analyses given above, $\sigma$ is independent of incident energy and centrality and also not evidently dependent on colliding system. The center parameter $y_0$ in (29) takes the values of 2.64–2.89, 2.63–2.83, 2.61–2.82, and 3.40–3.97 for centrality from small to large in Au-Au collisions at $\sqrt{s_{NN}} = 200$, 130, and 62.4 GeV and in Pb-Pb collisions at $\sqrt{s_{NN}} = 2.76$ TeV, respectively. As pointed out earlier, for a given colliding system, $y_0$ increases with energy and centrality.

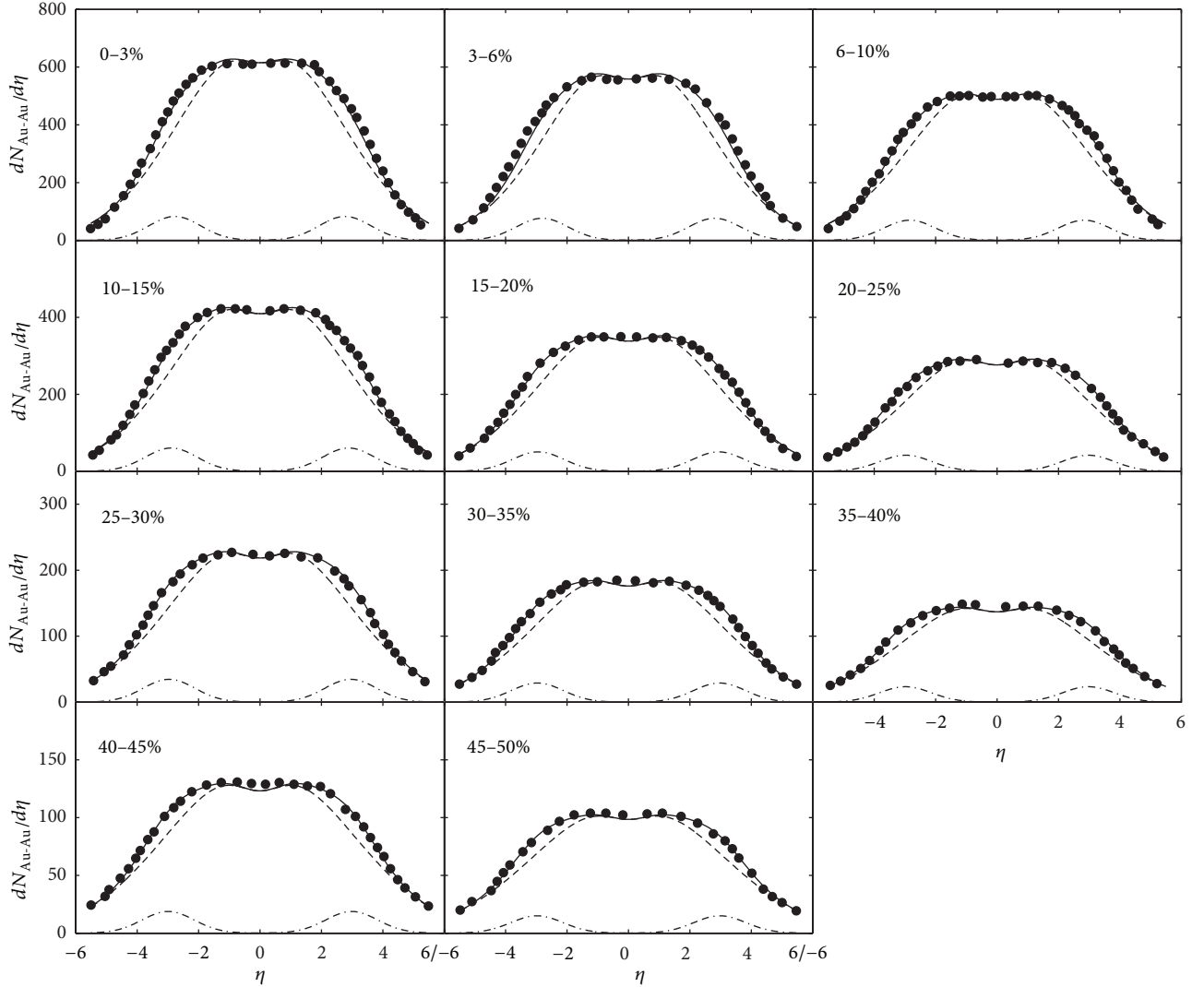

FIGURE 3: The pseudorapidity distributions of the charged particles produced in different centrality of Au-Au collisions at $\sqrt{s_{NN}} = 130$ GeV. The solid dots are the experimental measurements [36]. The dashed curves are the results from unified hydrodynamics of (28). The dashed-dotted curves are the results from leading particles of (29). The solid curves are the sums of the dashed and dashed-dotted ones.

## 5. Conclusions

By generalizing the relation between ordinary rapidity $y$ and space-time rapidity $\eta_{ST}$, unified hydrodynamics integrates the features of Hwa-Bjorken and Landau two famous hydrodynamic models together. In case of linear equation of state, this hydrodynamic model can be solved analytically. The exact solutions can then be used to formulate the rapidity distributions of the charged particles frozen out from the fluid at the space-like hypersurface with a fixed proper time $\tau_{FO}$. The unified hydrodynamics is successful in describing the rapidity distributions of $\pi^\pm$, $K^\pm$, and $\overline{p}$. However, it is not as good in describing the corresponding distributions of protons which might contain the leading particles.

In our previous work [6], we once discussed the same distributions in the context of evolution-dominated hydrodynamics. This hydrodynamic model differs from the one used in this paper in two ways. (1) Different initial conditions: the former assumes that the fluid is initially at rest. Its motion is totally dominated by the following evolution. The latter, as mentioned above, employs the initial condition of (7). It plays a connection between Hwa-Bjorken and Landau hydrodynamics. (2) Different freeze-out conditions:

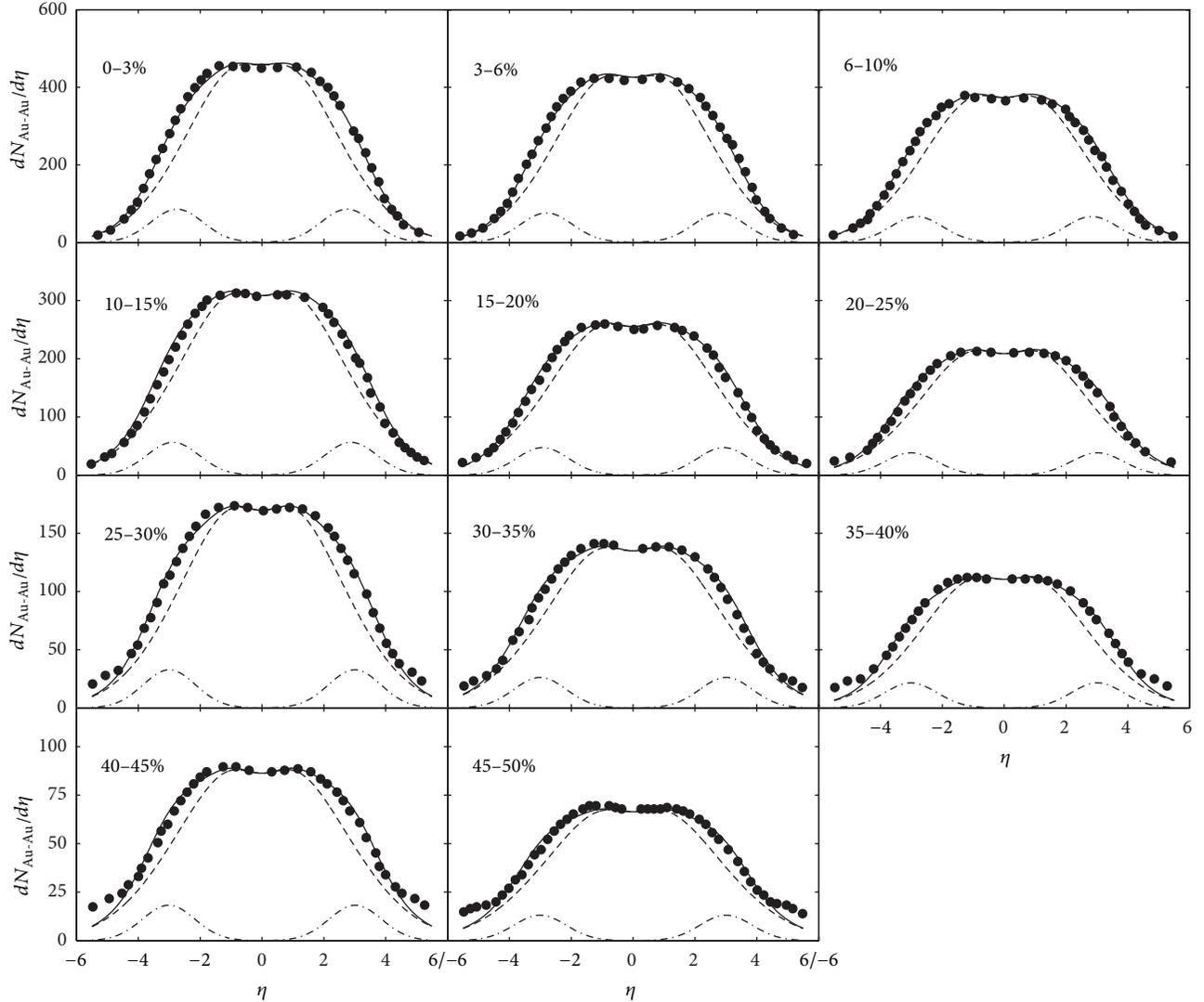

FIGURE 4: The pseudorapidity distributions of the charged particles produced in different centrality of Au-Au collisions at $\sqrt{s_{NN}}$ = 62.4 GeV. The solid dots are the experimental measurements [36]. The dashed curves are the results from unified hydrodynamics of (28). The dashed-dotted curves are the results from leading particles of (29). The solid curves are the sums of the dashed and dashed-dotted ones.

the former assumes that the freeze-out of fluid occurs at the space-like hypersurface with a fixed temperature $T_{FO}$. The latter, as stated above, takes such hypersurface as one with a fixed proper time $\tau_{FO}$.

As for leading particles, we argue that they possess the Gaussian rapidity distribution normalized to the number of participants. This is the same as that proposed in [6]. Here, for the purpose of completion and applications, we list out the most points of this proposition. It is interesting to notice that the investigations of the present paper once again show that, for a given colliding system, the central position $y_0$ of Gaussian rapidity distribution increases with incident energy and centrality, while the width $\sigma$ of the distribution is irrelevant to them and also almost independent of colliding system. These are consistent with the conclusions arrived at in [6].

Comparing with experimental data carried out by BRAHMS and PHOBOS Collaboration at BNL-RHIC in different centrality of Au-Au collisions at $\sqrt{s_{NN}}$ = 200, 130, and 62.4 GeV and by ALICE Collaboration at CERN-LHC in different centrality of Pb-Pb collisions at $\sqrt{s_{NN}}$ = 2.76 TeV, we can see that, although the charged particles frozen out from fluid play a dominant role, leading particles are also essential in characterizing the measured distributions. Only after the total contributions from both unified hydrodynamics and leading particles are taken into account together can the experimental measurements be matched up properly.

## Conflict of Interests

The authors declare that there is no conflict of interests regarding the publication of this paper.

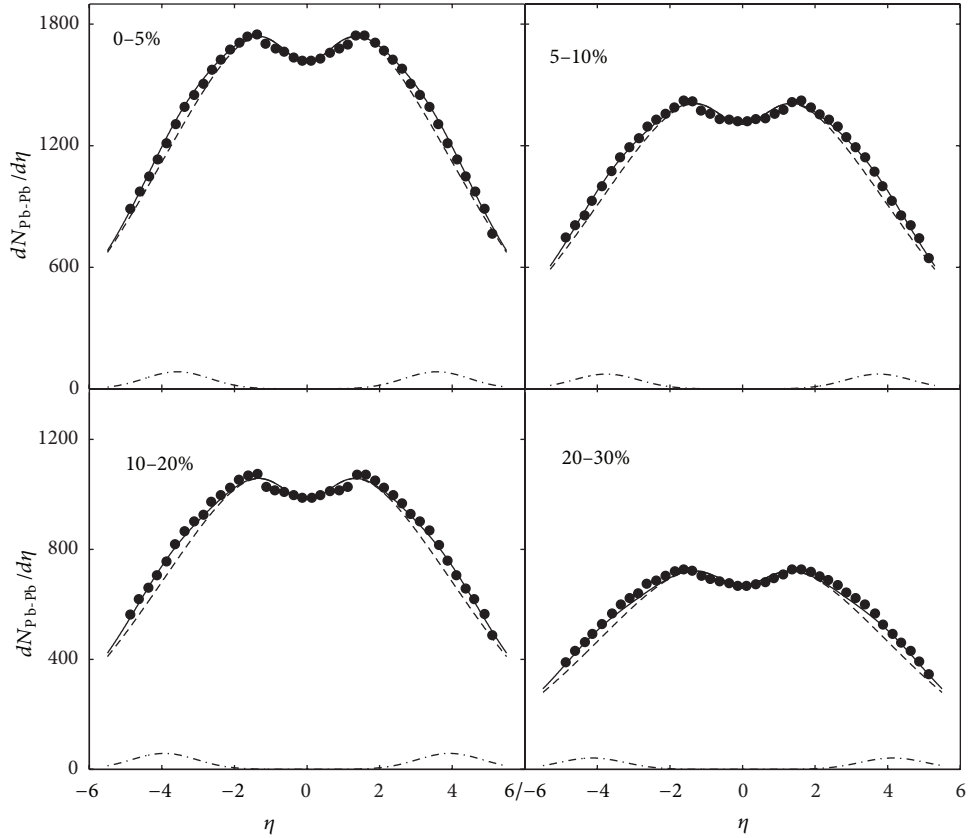

FIGURE 5: The pseudorapidity distributions of charged particles produced in different centrality of Pb-Pb collisions at $\sqrt{s_{NN}}$ = 2.76 TeV. The solid dots are the experimental measurements [37]. The dashed curves are the results from unified hydrodynamics of (28). The dashed-dotted curves are the results from leading particles of (29). The solid curves are the sums of the dashed and dashed-dotted ones.

## Acknowledgments

This work is partly supported by the Transformation Project of Science and Technology of Shanghai Baoshan District with Grant no. CXY-2012-25, the Shanghai Leading Academic Discipline Project with Grant no. XTKX 2012, the National Training Project with Grant no. 14XPM03, and the Hujiang Foundation of China with Grant no. B14004.

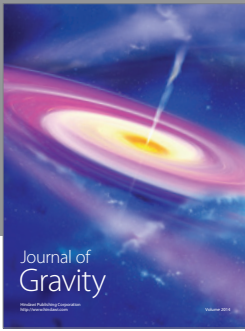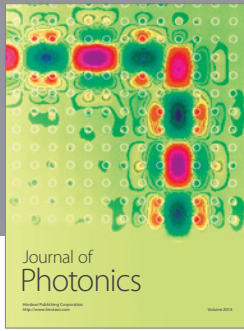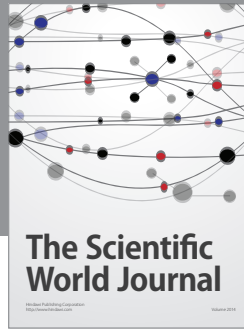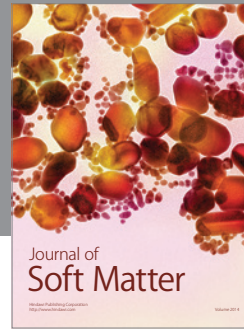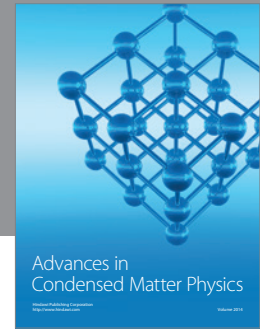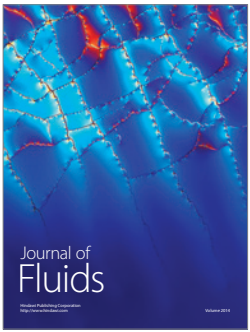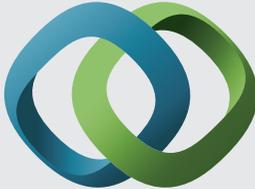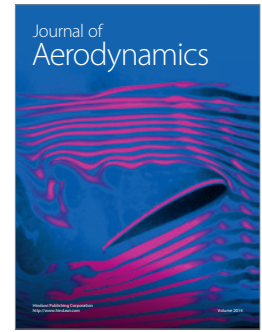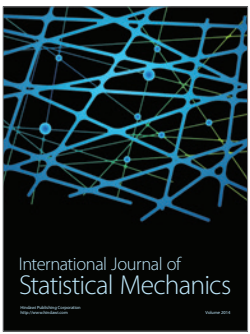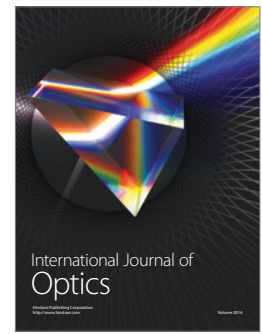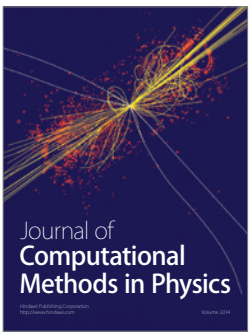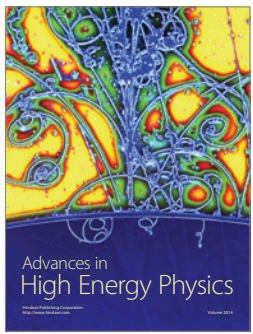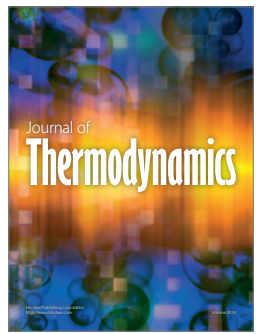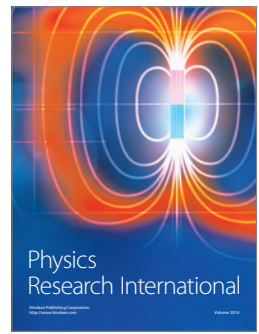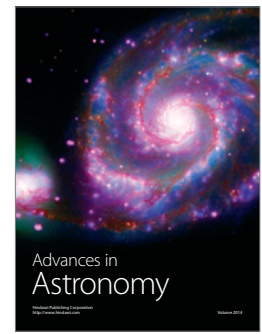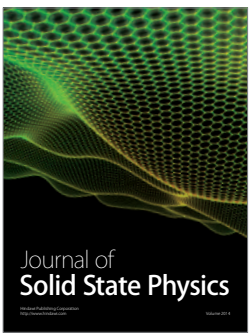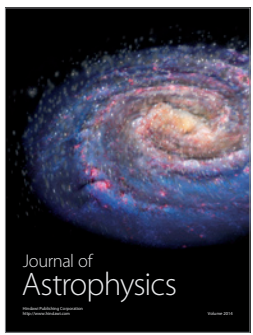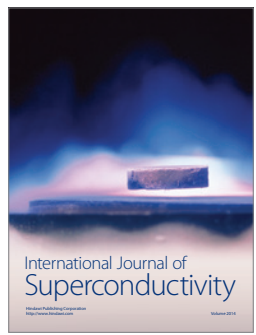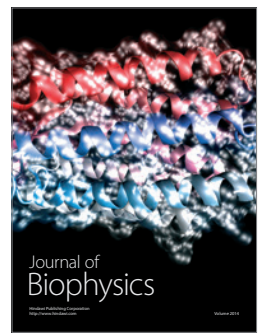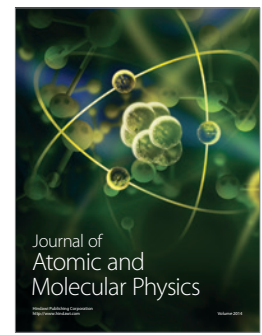